\documentstyle[12pt]{article}

\makeatletter
\def\thesection{\arabic{section}.}
\@addtoreset{equation}{section}

\def\appendix{\par
\setcounter{section}{0}
\setcounter{subsection}{0}
\def\thesection{\Alph{section}.}}
\makeatother
\makeatletter
\def\abstract#1{\long\def\@abstract{#1}}%
\def\@abstract{}%
\let\@oldmaketitle=\@maketitle%
\def\@maketitle{%
\@oldmaketitle%
\begin{center}\large\bf Abstract\end{center}%
\begin{quotation}\@abstract\end{quotation}%
\vskip 1.5em}%
\makeatother

\newcommand{\be}{\begin{equation} }
\newcommand{\ee}{\end{equation} }
\newcommand{\ba}{\begin{array}}
\newcommand{\ea}{\end{array}}
\newcommand{\beq}{\begin{eqnarray}}
\newcommand{\eeq}{\end{eqnarray}}
\newcommand{\lab} [1] {\label{eq:{#1}}}
\newcommand{\eqn} [1] {(~\ref{eq:{#1}}~)}
\newcommand{\bvphi}[1]{\varphi\hspace{-.617em}\varphi}
\newcommand{\bphi}[1]{\phi\hspace{-.58em}\phi}
\newcommand{\bfPhi}[1]{\Phi\hspace{-.74em}\Phi}
\newcommand{\bDelta}[1]{\Delta\hspace{-.79em}\Delta}

\begin{document}

\title{\sl A General Form of the Constraints in the Path Integral Formula}
\author{Taro KASHIWA\thanks{e-mail: taro1scp@mbox.nc.kyushu-u.ac.jp}\\
Department of Physics, Kyushu University, Fukuoka 812-81, Japan\\\\}

\date{May 26, 1995}
\abstract{A form of the constraints, specifying a $D$-dimensional manifold
embedded
in $D+1$ dimensional Euclidean space, is discussed in the path integral formula
given
by a time discretization. Although the mid-point prescription is privileged in
the
sphere $S^D$ case, it is more involved in generic cases. An interpretation on
the
validity of the formula is put in terms of the operator formalism. Operators
from
this path integral formula are also discussed. }
\maketitle\thispagestyle{empty}
\newpage

\section{Introduction}

Dynamical system, constrained on a $D$-dimensional manifold, $M^D$,
which is now supposed to be given by the equation,
\be
f({\bf x}) = 0,  \lab{1a}
\ee
where ${\bf x} \equiv \left(x^1, \cdots, x^{D+1}\right)$ is the
$D+1$-dimensional
Cartesian coordinate, can be described classically as follows: $f({\bf x})$ is
assumed to obey
\be
\left( \nabla_{\bf x} f({\bf x}) \right)^2 \neq 0;  \qquad  {}_{}^{\rm \rm
\forall }{\bf x}
\in M^D,  \lab{2a}
\ee
where we have written $\nabla_{\bf x}$ for the usual
$\nabla$ vector. The equation of motion in a flat $D+1$-dimensional space,
\be
\ddot{x}^a = - {\partial V({\bf x}) \over \partial x^a} \equiv - \partial_a
V({\bf
x}) ,
\lab{3a}
\ee
with $V({\bf x})$ being a potential, is modified to
\be
\Pi_{ab}(\nabla_{\bf x} f )\ddot{x}^b = - \Pi_{ab}(\nabla_{\bf x} f )\partial_b
V({\bf x}) ,
\lab{4a}
\ee
in $M^D$, where $\Pi_{ab}({\bf X} )$ is a projection operator,
\be
\Pi_{ab}({\bf X} ) \equiv \delta_{ab} - { X^a X^b \over {\bf X}^2 }, \lab{5a}
\ee
onto the plane perpendicular to the vector ${\bf X}$: $X^a\Pi_{ab}({\bf X}
)= \Pi_{ab}({\bf X} ) X^b = 0$. Here and
hereafter repeated indices imply summation. The significance of
\eqn{4a} is easily grasped; since the motion is restricted on $M^D$ so that any
deviation to the direction $\nabla_{\bf x} f$ must be suppressed.

It is well-known that the Lagrangian,
\be
L = {\dot{\bf x}^2 \over 2} - V({\bf x}) - \lambda f({\bf x}),  \lab{6a}
\ee
with $\lambda$ being the multiplier, leads to the equations \eqn{4a} and
\eqn{1a}.
Also the canonical formalism can be developed under the guidance of
Dirac\cite{Dirc}:
regard \eqn{1a} as the (primary) constraint
\be
\phi_1({\bf x}) \equiv f({\bf x})   \    \left( = 0; \qquad {}_{}^{\rm \rm
\forall }{\bf x}
\in M^D  \right) , \lab{7a}
\ee
and consider the consistency condition: a Hamiltonian,
\be
H = H({\bf p}, {\bf x} ) + \lambda f({\bf
x}) \equiv  { {\bf p}^2 \over 2} + V({\bf x}) + \lambda f({\bf
x}),
\lab{8a}
\ee
gives
\be
\dot{\phi}_1 = \{\phi_1, H\} = {\bf p}\cdot \nabla_{\bf x} f({\bf x}),
\lab{9a}
\ee
thus to find
\be
\phi_2({\bf x}) \equiv {\bf p}\cdot \nabla_{\bf x} f({\bf x})  \ \left( = 0;
\qquad
{}_{}^{\rm \rm \forall }{\bf x}
\in M^D  \right).
\lab{10a}
\ee
(Here $\{A, B\}$ designates the Poisson bracket.) They belong to the second
class:
\be
\{ \phi_1({\bf x}) , \phi_2({\bf x}) \} = \left( \nabla_{\bf x} f({\bf x})
\right)^2 \neq 0,
\lab{11a}
\ee
on account of \eqn{2a}, which enables us to obtain the Dirac bracket,
\be
\left\{A, B \right\}_{\rm D} \equiv \{A, B\} + { 1 \over  \left( \nabla_{\bf x}
f
\right)^2 }  \Big( \{A, \phi_1({\bf x}) \}  \{ \phi_2({\bf x}), B\} - ( A
\leftrightarrow B)
\Big).   \lab{12a}
\ee
Therefore we find
 \be
   \ba{l}
   {\displaystyle \left\{x^a, x^b \right\}_{\rm D} = 0 ,  }\\
  \noalign {\vskip 1ex}
{\displaystyle \left\{x^a, p_{\displaystyle {}_b} \right\}_{\rm D} =
\Pi_{ab}(\nabla_{\bf
x} f)=
\delta_{ab} - { \partial_a f \partial_b f \over \left( \nabla_{\bf x} f
\right)^2 } , }
\\
  \noalign {\vskip 1ex}
{\displaystyle  \left\{p_{\displaystyle {}_a}, p_{\displaystyle {}_b}
\right\}_{\rm D}=
p_{\displaystyle {}_c}
\left( \partial_a
\Pi_{c  b} -
\partial_b \Pi_{c a} \right)= p_{\displaystyle {}_c} {   \partial_a \partial_c
f
\partial_b f  -  \partial_b \partial_c  f
\partial_a  f \over
\left( \nabla_{\bf x} f \right)^2} ,}
 \ea
\lab{13a}
\ee
those which correctly reproduce the equation \eqn{4a}.

As for quantum mechanics, a recipe of path integral quantization had been given
by
Faddeev \cite{Fadd} and later by Senjanovic \cite{Senj}(FS): the FS-formula
reads
formally
\be
\langle \phi | e^{-iTH} | \psi \rangle = \int {\cal D} \mu \
 \phi^{*}({\bf x}_{f}) \exp \left[ i \int^{T/2}_{-T/2} dt \left\{ {\bf p} \cdot
\dot{\bf
x} - H({\bf p}, {\bf x} ) \right\} \right] \psi({\bf x}_i ) , \lab{14a}
\ee
with
\be
{\cal D} \mu \equiv {\cal D} {\bf p} {\cal D} {\bf x} \  | {\rm det} \{ \phi_1,
\phi_2 \}
|^{1/2} \delta (\phi_1 ) \delta ( \phi_2),    \lab{15a}
\ee
and ${\bf x}_f \equiv {\bf x}(T/2),  {\bf x}_i \equiv {\bf x}(-T/2).$ Here
${\cal D}
{\bf p}$ and ${\cal D} {\bf x}$ are functional measures which must be specified
somehow. The issue is then how to define the above functional measure properly
to
confirm the well-defined form of \eqn{14a}: the most well-known and primitive
approach is to discretize the time, obtaining
\be
{\cal D} {\bf p} \mapsto \prod_j d{\bf p}(j) ,  \quad {\cal D}  {\bf x} \mapsto
\prod_j d{\bf x}(j).  \lab{16a}
\ee
In this approach it was stressed by the present author \cite{FK} that the
mid-point
prescription is privileged in the case of $D$-dimensional sphere $S^D$ given as
\be
{\bf x}^2 = \rho^2.  \lab{17a}
\ee
We try to generalize the case in this paper.

In section 2, we review the $S^D$ case. With this in mind, a generic case
$f({\bf
x}) =0$ is discussed in section 3. The next section 4 deals with operators
obtained from the path integral formula, then the final section 5 is devoted to
discussion.

\section{The case of $D$-dimensional sphere}

The $D$-dimensional sphere $S^D$ is given, in view of \eqn{17a}, by
\be
f({\bf x}) \equiv {1 \over 2} \left( {\bf x}^2 - \rho^2 \right) ( = \phi_1).
\lab{1b}
\ee
The secondary constraint \eqn{10a} is read as
\be
\phi_2 \equiv {\bf p} \cdot \nabla_{\bf x} f({\bf x}) = {\bf p} \cdot {\bf x}.
\lab{2b}
\ee

The FS-formula \eqn{14a} and \eqn{15a} in a discretized form is found as
\be
\ba{l}
{\displaystyle \langle \phi | e^{-iTH} | \psi \rangle \equiv \lim_{N
\rightarrow
\infty} \prod_{j=0}^{N} \int d^{D+1} {\bf x}(j) \  \delta \! \left( \phi_1\!
\left({\bf
x}(j)
\right) \right)  }\\
\noalign {\vskip 1ex}
{\displaystyle \hspace{10ex} \times \prod_{j=1}^{N} \int { d^{D+1} {\bf p}(j)
\over (2
\pi )^D }
 \ \delta \! \left( \phi_2(j) \right)   | {\rm det} \left\{ \phi_1\! \left({\bf
x}(j)
\right) , \phi_2(j) \right\} |^{1/2}  }\\
\noalign {\vskip 1ex}
{\displaystyle \times  \phi^{*}({\bf x}(N)) \exp \left[ i \sum_{j=1}^{N}
\bigg\{
{\bf p} (j) \cdot \Delta {\bf x}(j) - \Delta t  H({\bf p}(j), \overline{\bf
x}(j) ) \bigg\}
\right] \psi({\bf x}(0) ) ,}
\ea \lab{3b}
\ee
with
\be
\Delta t \equiv { T \over N},  \lab{4b}
\ee
\be
\Delta {\bf x}(j) \equiv {\bf x}(j) - {\bf x}(j-1)  , \lab{5b}
\ee
and
\be
\overline{\bf x}(j) \equiv { {\bf x}(j) + {\bf x}(j-1) \over 2}. \lab{6b}
\ee
Here we have employed the mid-point prescription \eqn{6b} to the argument of
Hamiltonian, which can be interpreted as a consequence of the Weyl
ordering \cite{Coh}\cite{K}. The issue is to fix the form of $\phi_2(j)$: the
correct form
has been found also as the mid-point type \cite{FK}:
\be
\phi_2(j) = {\bf p}(j) \cdot \overline{\bf x}(j). \lab{7b}
\ee
The way to \eqn{7b} can be convinced by the following discussion.

Consider $T=0$ case: put $N=1$ in \eqn{3b} to obtain
\be
\ba{c}
{\displaystyle \langle \phi | \psi \rangle = \int d^{D+1} {\bf x} \  d^{D+1}
{\bf x'}
\ \delta
\! \left( { {\bf x}^2- \rho^2 \over 2}  \right)  \delta \! \left( { {\bf x'}^2-
\rho^2 \over 2}
\right)}\\
\noalign {\vskip 1ex}
{\displaystyle \times \int { d^{D+1} {\bf p} \over (2 \pi )^D } \
\delta \! \left({\bf p} \cdot {\bf x}^{(\alpha)} \right)  \left| {\bf x} \cdot
{\bf
x}^{(\alpha)} \right|  \phi^{*}({\bf x}) e^{i {\bf p} \cdot ( {\bf x} - {\bf
x'} )}
\psi({\bf x'} ), }
\ea \lab{8b}
\ee
where we have written ${\bf x}, {\bf x'},$ and ${\bf p}$ for ${\bf x}(1), {\bf
x}(0),$ and
${\bf p}(1)$ respectively and set the form of \eqn{2b} as
\be
\phi_2(j=1) =  {\bf p} \cdot {\bf x}^{(\alpha)} \equiv {\bf p} \cdot \left( ({1
\over2 } - \alpha ) {\bf x} + ({1 \over2 } +
\alpha ) {\bf x'}\right)  , \lab{9b}
\ee
with $\alpha$ being a parameter \cite{K} to be determined. Decompose
the ${\bf p}$-vector such that
\be
{\bf p} = {\bf p}_{\!{}_{/ \! /}} + {\bf p}_{\!{}_{ \perp}} , \lab{10b}
\ee
where
\be
\ba{c}
{\displaystyle  {\bf p}_{\!{}_{/ \! /}} \equiv { {\bf p} \cdot {\bf
x}^{(\alpha)} \over
 {  {\bf x}^{(\alpha)} }^2 } {\bf x}^{(\alpha)}, }\\
\noalign {\vskip 1ex}
{\displaystyle \left( {\bf p}_{\!{}_{
\perp}}\right)_a
\equiv
\Pi_{ab}  \!\left({\bf x}^{(\alpha)} \right)  p_{\displaystyle {}_b},   }
\ea  \lab{11b}
\ee
are the parallel and the perpendicular components to the vector ${\bf
x}^{(\alpha)}$. Then perform the
${\bf p}$-integration to find
\be
\int { d^{D+1} {\bf p} \over (2 \pi )^D } \
\delta \! \left({\bf p} \cdot {\bf x}^{(\alpha)} \right)  e^{i {\bf p} \cdot (
{\bf x} -
{\bf x'} )} = { 1 \over | {\bf x}^{(\alpha)} | } \delta^D \! \Big( ( {\bf x} -
{\bf x'} )_{\!{}_{\perp}} \Big), \lab{12b}
\ee
where
\be
 ( {\bf x} - {\bf x'} )_{\!{}_{\perp}}^a  \equiv  \Pi_{ab}  \!\left({\bf
x}^{(\alpha)} \right) \left( {\bf x} - {\bf x'}  \right)^b.  \lab{13b}
\ee
Therefore the $D$-dimensional $\delta$-function, in the right hand side of
\eqn{12b}, implies
\be
0= ( {\bf x} - {\bf x'} )_{\!{}_{\perp}}^a = \left( {\bf x} - {\bf x'}
\right)^a - { {\bf
x}^{(\alpha)} \cdot \left( {\bf x} - {\bf x'}  \right) \over ({\bf
x}^{(\alpha)} )^2 }
\left( {\bf x}^{(\alpha)} \right)^a,  \lab{14b}
\ee
with the aid of  \eqn{5a}. The solution is
\be
{\bf x} = {\bf x'}, \qquad \mbox{ for $\alpha =0$,}  \lab{15b}
\ee
since the second term of \eqn{14b} vanishes:
\be
 {\bf x}^{(\alpha=0)} \cdot \left( {\bf x} - {\bf x'}  \right) = {1 \over 2}
\left( {\bf x}^2 -
{\bf x'}^2 \right) = 0,  \lab{16b}
\ee
owing to the constraint \eqn{1b}. But an additional point emerges if
$\alpha \neq 0$
\be
{\bf x} = - {\bf x'}.   \lab{17b}
\ee
Thus in $\alpha \neq 0$ the
$\delta$-function in \eqn{12b} is double-valued. To avoid the situation we must
take
$\alpha =0$, that is,
\eqn{9b} turns out to be \eqn{7b}.

\section{A path integral formula in generic cases}

In this section we wish to generalize the previous result to $M^D$, given by
$f({\bf x}) =0$. Start from \eqn{3b} by putting
\be
\phi_2(j) \equiv {\bf p}(j) \cdot \nabla \! f(j), \lab{1c}
\ee
and study the form of $\nabla \! f(j)$. The ${\bf p}(j)$-integral in this case
becomes
\be
\int { d^{D+1} {\bf p}(j) \over (2 \pi )^D } \
\delta \! \left({\bf p}(j) \cdot \nabla \! f(j) \right)  e^{i {\bf p}(j) \cdot
\Delta {\bf
x}(j) } = { 1 \over | \nabla \! f(j) | } \delta^D \! \left( \Delta {\bf
x}_{\!{}_{\perp}}(j)
\right), \lab{2c}
\ee
where
\be
\Delta {\bf x}_{\!{}_{\perp}}^a(j) = \Pi_{ab}(\nabla \! f(j) ) \Delta x^b(j) =
\Delta
x^a(j) - { \Delta {\bf x}(j) \cdot  \nabla \! f(j) \over \nabla \! f(j)^2
} \big( \nabla \! f(j)
\big)^a
  ,  \lab{3c}
\ee
which is again the consequence of the decomposition of ${\bf p}$'s into the
parallel
and the perpendicular components with respect to a (still unknown) vector
$\nabla \!
f(j)$.

According to the foregoing discussion, \eqn{14b} $\sim$ \eqn{17b},  a
sufficient
condition for a single-valued $\delta$-function on $M^D$ is read from \eqn{3c}
\be
\Delta {\bf x}(j) \cdot  \nabla \! f(j) =0;  \qquad  {}_{}^{\rm \rm \forall
}{\bf x}
\in M^D.  \lab{4c}
\ee
A simple solution therefore is
\be
\Delta {\bf x}(j) \cdot  \nabla \! f(j)= f\big({\bf x}(j) \big) - f\big({\bf
x}(j-1) \big).
\lab{5c}
\ee
(This would make sense; since a naive continuum limit, defined by ${\bf x}(j),
{\bf
p}(j) \rightarrow {\bf x}(t), {\bf p}(t)$, ${\bf x}(j-1) \rightarrow {\bf
x}(t-dt)$, implies
$\nabla
\! f(j)
\rightarrow \nabla_{\bf x} f({\bf x} )$, yielding the classical result
\eqn{10a}. )
Write
\be
\ba{c}
{\displaystyle {\bf x}(j) = \overline{\bf x}(j)+ { \Delta {\bf x}(j) \over 2} ,
 }\\
\noalign {\vskip 1ex}
{\displaystyle {\bf x}(j-1) = \overline{\bf x}(j) - { \Delta {\bf x}(j) \over
2}  ,}
\ea \lab{6c}
\ee
and expand the right hand side of \eqn{5c} with respect to $\Delta {\bf x}(j)$
to
obtain
\be
\nabla \! f(j) =  \left\{ \sum_{n=0}^{\infty} { 1 \over ( 2n
+1 )! }
\left( { \Delta {\bf x}(j) \cdot \nabla_{\overline{\bf x}} \over 2}
\right)^{2n} \right\}
\nabla_{\overline{\bf x}} f\!\left(
\overline{\bf x}(j) \right) ,  \lab{7c}
\ee
where $\nabla_{\overline{\bf x}}$ denotes differentiation with respect to
$\overline{\bf x}(j)$. With this in mind a path integral formula on $M^D$ is
found as
\be
\ba{c}
{\displaystyle \langle \phi | e^{-iTH} | \psi \rangle \equiv \lim_{N
\rightarrow
\infty} \prod_{j=0}^{N} \int d^{D+1} {\bf x}(j)  \  \delta \! \left( f\!
\left({\bf x}(j)
\right) \right) }\\
\noalign {\vskip 1ex}
{\displaystyle \times \prod_{j=1}^{N} \int { d^{D+1} {\bf p}(j) \over (2 \pi
)^D } \
\delta \! \left( {\bf p}(j) \cdot \nabla \! f(j) \right) |  \nabla_{\bf x} f\!
\left({\bf
x}(j)
\right)  \cdot \nabla \! f(j) |  }\\
\noalign {\vskip 1ex}
{\displaystyle \times  \phi^{*}({\bf x}(N)) \exp \left[ i \sum_{j=1}^{N}
\left\{
{\bf p} (j) \cdot \Delta {\bf x}(j) - \Delta t  H({\bf p}(j), \overline{\bf
x}(j) ) \right\}
\right] \psi({\bf x}(0) ).}
\ea \lab{8c}
\ee

Needless to say,  \eqn{1c} with \eqn{7c} matches \eqn{7b}, the $S^D$ case,
where
symmetry is higher so that the mid-point prescription was valid. But as can be
recognized from \eqn{7c} {\em there is no privilege of the mid-point
prescription in
general cases.}

Before closing this section let us argue another aspect of the relation
\eqn{2c} with
\eqn{7c}: on $M^D$, $x^a$ can be expressed by
some coordinate, say, $\theta^i$  $(i=1, 2, \cdots, D)$:
\be
x^a = x^a( \theta), \qquad \qquad \theta \in M^D.  \lab{9c}
\ee
There should be an orthonormal as well as complete set, $Y_n(\theta)$:
\be
\int  d^D \theta  \sqrt{ g(\theta) } Y_n^*(\theta) Y_{n'} (\theta) = \delta_{n,
n'},
\lab{10c}
\ee
\be
\sum_n Y_n(\theta) Y_n^* (\theta') = { 1 \over \sqrt{ g(\theta) } } \delta^D(
\theta -
\theta'),  \lab{11c}
\ee
where $n$ represents generic labels and $g(\theta)$ is the determinant of the
induced
metric,
\be
g_{ij}(\theta) = \sum_{a=1}^{D+1} { \partial x^a \over \partial \theta^i} {
\partial x^a
\over \partial \theta^j}. \lab{12c}
\ee
Specifically, $Y_n( \theta)$ may be an eigenfunction of the Laplace-Beltrami
operator:
\be
- \left[ g^{-1/2} { \partial \over \partial \theta^i } \left( g^{ij} g^{ 1/2}
\right) { \partial
\over \partial \theta^j } \right] Y_n( \theta) = h(n ) Y_n(\theta).  \lab{13c}
\ee

Suppose that Hamiltonian is given by
\be
\hat{H}= -   g^{-1/2} { \partial \over \partial \theta^i } \left( g^{ij} g^{
1/2} \right) {
\partial \over \partial \theta^j }  + V(\hat{\theta}), \lab{14c}
\ee
where the caret denotes operators, then the Feynman kernel,
\be
K( \theta, \theta'; T) \equiv \langle \theta | e^{-iT\hat{H}} | \theta' \rangle
= \lim_{N
\rightarrow \infty} \langle \theta | \left( {\bf I} -i \Delta t \hat{H}
\right)^N | \theta'
\rangle ,
\lab{15c}
\ee
can be expressed as ``path integral'': by inserting the identities, \eqn{10c}
and
\eqn{11c}, which are now read as
\be
 \int d^D \theta \ \sqrt{ g(\theta) } \  | \theta \rangle \langle \theta | =
{\bf I},
\lab{16c}
\ee
\be
 \sum_n | n \rangle \langle n | = {\bf I}, \lab{17c}
\ee
with ${\bf I}$ being the identity operator,
\be
\ba{l}
{\displaystyle \langle \theta | \theta' \rangle = { 1 \over \sqrt{ g(\theta) }
} \delta^D
(\theta -
\theta'), } \\
\noalign{\vskip 1ex}
{\displaystyle \langle n | n' \rangle = \delta_{n n'}, }
\ea   \lab{18c}
\ee
and $\langle \theta | n \rangle \equiv Y_n(\theta)$, \eqn{15c} becomes
\be
\ba{l}
{\displaystyle  K( \theta, \theta'; T) = \lim_{N \rightarrow \infty}
\left(  \prod_{j=1}^{N-1}  \int d^D \theta(j) \ \sqrt{ g \Big( \theta(j) \Big)
} \right)
\Bigg(
\prod_{j=1}^{N} \sum_{n(j)}   }  \\
\noalign {\vskip 1ex}
{\displaystyle  \qquad  \qquad   \times  Y_{n(j) }\! \left( \theta(j) \right)
Y_{n(j) }^*\!  \left( \theta(j-1) \right) \exp[ -i \Delta t
\left\{  h( n(j) )  + V\! \left(\theta(j) \right)  \right\}  ]  \Bigg)
\Bigg|^{\theta(N)
= \theta}_{\theta(0) = \theta'}.  }
\ea  \lab{20c}
\ee
However the expression of \eqn{20c} is unsatisfactory as a ``path integral''
formula if
$M^D$ is nontrivial, $g_{ij} \neq \delta_{ij}$; since some of the labels are
discrete so that
we are left with summation not integration. Moreover
$Y_n(\theta)$ is generally far from a plane wave form: in a trivial case,
$g_{ij}=
\delta_{ij}$, (which is given by an $f({\bf x})$ linear in ${\bf x}$,)
$Y_n(\theta)$ is read as,
\be
Y_n(\theta) \equiv { 1 \over ( 2 \pi )^{D/2} } e^{ i {\bf P} \cdot {\bf X} }.
\lab{21c}
\ee
($n$ and $\theta$ correspond to ${\bf P}$ and ${\bf X}$ respectively.)
Therefore we
obtain a usual path integral formula:
\be
\ba{l}
{\displaystyle K\!\left( {\bf X}, {\bf X'}; T \right) = \lim_{N \rightarrow
\infty}
\left(  \prod_{j=1}^{N-1}  \int d^D {\bf X}(j) \right)
\left(
\prod_{j=1}^{N} \int { d^D {\bf P}(j) \over (2 \pi )^D } \right)}  \\
\noalign {\vskip 1ex}
{\displaystyle  \     \times   \exp \! \!  \left.\left[ i \sum_{j=1}^{N}
\Bigg\{ {\bf P}(j) \cdot
\Delta {\bf X}(j)  - \Delta t
\Big(  h\!\left( {\bf P}(j) \right)  + V\! \left({\bf X}(j) \right)  \Big)
\Bigg\} \right] \!
\right|^{{\bf X}(N) = {\bf X}}_{{\bf X}(0) = {\bf X}^{'}}.  }
\ea  \lab{22c}
\ee
(It might be natural, however, to think that the situation is same even in the
trivial
case if we work with the polar coordinate; since in which there arises the
spherical
harmonics, being far from the plane wave  except the $S^1$ case. But in these
cases we
can find a desired path integral formula consisting purely of an exponential
form as
well as integration by means of the canonical transformation \cite{FK2} from
the
Cartesian expression \eqn{22c}.)

Now it is almost clear that the relation \eqn{2c} with \eqn{7c} cures the above
situation
for nontrivial cases:  according to our discussion, the completeness condition
\eqn{11c}
can be put into a plane wave type provided solely with integration:
\be
\sum_n Y_n(\theta) Y_n^* (\theta') = { 1 \over \sqrt{ g(\theta) } } \delta^D(
\theta -
\theta') = | \nabla f | \int { d^{D+1} {\bf p} \over (2 \pi )^D } \ \delta\!
\left( {\bf p}
\cdot \nabla f \right) e^{ i {\bf p} \cdot ( {\bf x} - {\bf x'} ) } \Bigg|_{M},
\lab{23c}
\ee
where from \eqn{7c}
\be
\nabla f \equiv \left\{ \sum_{n=0}^\infty { 1 \over (2n+1)! } \left( { \Delta
{\bf x}
\cdot \nabla_{\overline{\bf x} } \over 2}  \right)^{2n}  \right\}
\nabla_{\overline{\bf x} } f\!\left(\overline{\bf x} \right), \lab{24c}
\ee
and the subscript $M$ designates that ${\bf x}$ and ${\bf x'}$ are on $M^D$.
The relation \eqn{23c} thus can be implied as {\em the plane wave
representation of
the completeness condition} on
$M^D$. In other words the FS formula is a rigorous
consequence from the operator formalism owing to this completeness condition
\eqn{23c}.

\section{Operators from the path integral formula}

A similar consideration as in \eqn{8b} leads us to the observation that an
expectation
value of some operator ${\cal O}\!\left(\hat{\bf p}, \hat{\bf x}\right)$ can be
given,
with the aid of the formula \eqn{8c} with \eqn{7c}, by
\be
\ba{l}
{\displaystyle  \langle {\cal O}\!\left(\hat{\bf p}, \hat{\bf x}\right) \rangle
\equiv
\langle \varphi | {\cal O}\!\left(\hat{\bf p}, \hat{\bf x}\right) | \psi
\rangle \equiv
\int d^{D+1} {\bf x} \int d^{D+1} {\bf x'}  \  \delta \!\left( f({\bf x} )
\right) \delta
\!\left( f({\bf x'} ) \right) }\\
\noalign {\vskip 1ex}
{\displaystyle   \times  \left| \nabla_{\bf x} f({\bf x}) \cdot
\nabla f \right|  \varphi^*({\bf x}) \psi({\bf x'} )\int { d^{D+1} {\bf p}
\over (2
\pi)^D }  \ \delta\! \left( {\bf p}\cdot \nabla f \right) {\cal O}\!\left( {\bf
p},
\overline{\bf x} \right) e^{i {\bf p}\cdot \Delta {\bf x} },  }
\ea \lab{1d}
\ee
where $\nabla f$ is given by \eqn{24c},
\be
\Delta {\bf x} \equiv {\bf x} - {\bf x'}, \lab{2d}
\ee
and
\be
\overline{\bf x} \equiv { {\bf x} + {\bf x'} \over 2} . \lab{3d}
\ee
By noting
\be
\delta(X) \delta(Y) = \delta\!\left( { X+Y \over 2} \right) \delta(X-Y),
\lab{4d}
\ee
then using
\eqn{5c}, \eqn{1d} becomes
\be
\ba{c}
{\displaystyle  \langle {\cal O}\!\left(\hat{\bf p}, \hat{\bf x}\right) \rangle
  =  \int
d^{D+1} {\bf x} \int d^{D+1} {\bf x'}  \  \delta \!\left( \overline{f}  \right)
\delta\!\left( \Delta {\bf x} \cdot \nabla f \right) \left|
\nabla_{\bf x} f({\bf x}) \cdot \nabla f \right| \varphi^*({\bf x}) \psi({\bf
x'} ) }\\
\noalign {\vskip 1ex}
{\displaystyle \times \int { d^{D+1} {\bf p} \over (2
\pi)^D }  \ \delta\! \left( {\bf p}\cdot \nabla f \right) {\cal O}\!\left( {\bf
p},
\overline{\bf x} \right) e^{i {\bf p}\cdot \Delta {\bf x} = \int d^{D+1} {\bf
x} \int
d^{D+1} {\bf x'}  }}\\
\noalign {\vskip 1ex}
{\displaystyle  \times  \delta \!\left( \overline{f}
\right)  { \left| \nabla_{\bf x} f({\bf x}) \cdot
\nabla f \right| \over \left( \nabla f \right)^2 }\varphi^*({\bf x})
\psi({\bf x'} )  {\cal O}\!\left( -i { \partial \over \partial \Delta {\bf
x}_{\perp} }, \overline{\bf x} \right) \delta^{D+1}\!\left( \Delta {\bf x}
\right) , }
\ea \lab{5d}
\ee
where we have introduced the notation,
\be
\overline{f} \equiv {  f({\bf x} ) +  f({\bf x'} ) \over 2 }, \lab{6d}
\ee
and integrated with respect to ${\bf p}$'s in a similar manner as before, to
find
$\delta^D\!\left( \Delta {\bf x}_{\perp} \right)$ which is combined with
$\delta\!
\left( \Delta {\bf x} \cdot \nabla f \right) \sim \delta\! \left( \Delta {\bf
x}_{\!{}_{/
\! /}}
\right)$ yielding $\delta^{D+1}(\Delta {\bf x} )$ finally. Now changing
variables $(
{\bf x}, {\bf x'} )$ to $(\overline{\bf x}, \Delta {\bf x})$ and performing
integration
by parts, we find
\be
\ba{l}
{\displaystyle  \langle {\cal O}\!\left(\hat{\bf p}, \hat{\bf x}\right) \rangle
 =   \int
d^{D+1} \overline{\bf x}  \  {\cal O}\!\left( -i { \partial
\over \partial \Delta {\bf x}_{\perp} }, \overline{\bf x} \right)  }\\
\noalign {\vskip 1ex}
{\displaystyle \hspace{7ex}  \left. \times \left[ \delta \!\left(
\overline{f}  \right) { \left| \nabla_{\bf x} f({\bf x}) \cdot
\nabla f \right| \over \left( \nabla f \right)^2 }  \varphi^*\!
\left(\overline{\bf x} +
{\Delta {\bf x} \over 2} \right)
\psi\! \left(\overline{\bf x} - {\Delta {\bf x} \over 2} \right) \right]
\right|_{\Delta {\bf x}=0 } , }
\ea \lab{7d}
\ee
where
\be
\nabla_{\bf x} f({\bf x}) = \left\{ \sum_{n=0}^\infty { 1 \over n! } \left( {
\Delta {\bf
x} \cdot \nabla_{\overline{\bf x} } \over 2} \right)^n \right\}
\nabla_{\overline{\bf
x} } f\!\left(\overline{\bf x} \right),  \lab{8d}
\ee
and
\be
\overline{f} = \sum_{n=0}^\infty { 1 \over (2n)! } \left( { \Delta {\bf
x} \cdot \nabla_{\overline{\bf x} } \over 2} \right)^{2n}
f\!\left(\overline{\bf x}
\right), \lab{9d}
\ee
in view of \eqn{6d}. (The subscript $\Delta {\bf x} =0$ designates that $\Delta
{\bf
x} \rightarrow 0$ must be put after all calculations have been done.) Also note
that
\be
\ba{c}
{\displaystyle { \left| \nabla_{\bf x} f({\bf x}) \cdot
\nabla f \right| \over \left( \nabla f \right)^2 }= 1 + { 1 \over
\left( \nabla_{\overline{\bf x}  }  f\!\left(\overline{\bf x} \right) \right)^2
}
\nabla_{\overline{\bf x}  }  f\!\left(\overline{\bf x} \right) \cdot
\nabla_{\overline{\bf x}  }\left( { \Delta {\bf x} \cdot \nabla_{\overline{\bf
x} }
\over 2} \right) f\!\left(\overline{\bf x} \right)  }\\
\noalign {\vskip 1ex}
{\displaystyle + { 1 \over
3 \left( \nabla_{\overline{\bf x}  }  f\!\left(\overline{\bf x} \right)
\right)^2 }
\nabla_{\overline{\bf x}  }  f\!\left(\overline{\bf x} \right) \cdot
\nabla_{\overline{\bf x}  }\left( { \Delta {\bf x} \cdot \nabla_{\overline{\bf
x} }
\over 2} \right)^2 f\!\left(\overline{\bf x} \right)   + O\!\left(\Delta {\bf
x}^3\right). }
\ea \lab{10d}
\ee

Let us calculate some examples:
\begin{itemize}
\item (i) ${\cal O}\left(\hat{\bf p}, \hat{\bf x}\right) \equiv F\!\left(
\hat{\bf x}
\right)$:
\end{itemize}
\be
\langle F\!\left( \hat{\bf x} \right) \rangle = \int d^{D+1} {\bf x}  \
\delta
\!\left( f({\bf x} ) \right)  \varphi^*\! \left({\bf x}\right) F\! \left({\bf
x}\right)
\psi\! \left( {\bf x}  \right) ,  \lab{11d}
\ee
where we have written ${\bf x}$ for $\overline{\bf x}$. This shows
\be
F\!\left( \hat{\bf x} \right) =  F\! \left({\bf x}\right) .  \lab{12d}
\ee
\begin{itemize}

\item(ii) ${\cal O}\left(\hat{\bf p}, \hat{\bf x}\right) \equiv
\hat{p}_{\displaystyle {}_a} $:

\end{itemize}

\be
\ba{l}
{\displaystyle  \langle \hat{p}_{\displaystyle {}_a} \rangle  =   \int
d^{D+1} \overline{\bf x}  \   \Pi_{ab}\!\left(\nabla_{\overline{\bf x}} f
\right)   \
\left( -i { \partial \over \partial \Delta x^b } \right) \Bigg[ \delta \!\left(
\overline{f}  \right)  { \left| \nabla_{\bf x} f({\bf x}) \cdot
\nabla f \right| \over \left( \nabla f \right)^2 }  }\\
\noalign {\vskip 1ex}
{\displaystyle    \times   \varphi^*\! \left(\overline{\bf x} +
{\Delta {\bf x} \over 2} \right)
\psi\! \left(\overline{\bf x} - {\Delta {\bf x} \over 2} \right) \Bigg]
\Bigg|_{\Delta {\bf x}=0 }  = \int d^{D+1} {\bf x} \  \delta \!\left( f ({\bf
x} ) \right) }\\
\noalign {\vskip 1ex}
{\displaystyle  \times
\Pi_{ab}\!\left(\nabla_{\bf x } f \right) {i \over 2} \Bigg\{  \partial_b
\varphi^*\! \left({\bf x} \right)  \psi\! \left( {\bf x} \right)  -
\varphi^*\! \left({\bf
x} \right) \partial_b \psi\! \left( {\bf x} \right) + { \partial_b\partial_c f
\partial_c f \over \left(\nabla_{\bf x } f\right)^2}  \varphi^*\! \left({\bf
x} \right) \partial_b \psi\! \left( {\bf x} \right) \Bigg\} , }
 \ea
\lab{13d}
\ee
where again we have put $\overline{\bf x} \rightarrow {\bf x}$. The third term
in
the final expression comes from the differentiation to \eqn{10d}. (There
remains no
effect from differentiating the $\delta$-function, in view of
\eqn{9d}.)  Finally integrating by parts with respect to the first term, while
paying attention to the property of the projection operator, $\Pi_{ab}
\partial_b
\delta \! \left( f({\bf x}) \right) =0$, we obtain
\be
\ba{l}
{\displaystyle  \langle \hat{p}_{\displaystyle {}_a} \rangle  =   \int d^{D+1}
{\bf x}  \
  \delta \!\left( f ({\bf x} ) \right)  \  \varphi^*\! \left({\bf x}
\right)  \bigg[ -i \Pi_{ab}\!\left(\nabla_{\bf x } f \right) \partial_b }\\
\noalign {\vskip 1ex}
{\displaystyle  -{ i \over 2}  \partial_b \Pi_{ab}\!\left(\nabla_{\bf x } f
\right)
-{ i \over 2}  \Pi_{ab}\!\left(\nabla_{\bf x } f \right) \partial_c
\Pi_{bc}\!\left(\nabla_{\bf x } f \right) \bigg]   \psi\! \left( {\bf x}
\right). }
\ea \lab{14d}
\ee
Therefore
\be
\ba{l}
{\displaystyle  \hat{p}_{\displaystyle {}_a}  =   -i
\Pi_{ab}\!\left(\nabla_{\bf x } f
\right) \partial_b
 -{ i \over 2}  \partial_b \Pi_{ab}\! \left(\nabla_{\bf x } f \right)  -{ i
\over 2}  \Pi_{ab}\!\left(\nabla_{\bf x } f \right)
\partial_c
\Pi_{bc}\!\left(\nabla_{\bf x } f \right) }\\
\noalign {\vskip 1ex}
{\displaystyle = -i \Pi_{ab}\!\left(\nabla_{\bf x } f \right) \partial_b + { i
\over 2}
{ 2 \partial_a \partial_b f \partial_b f  + \partial_a f {\nabla_{\bf x }}^2 f
\over
\left( \nabla_{\bf x } f \right)^2 }  -  {3 i \over 2}
{  \partial_b \partial_c f  \partial_a f \partial_b f \partial_c f   \over
\left( \nabla_{\bf x } f \right)^4 } ,  }
\ea \lab{15d}
\ee
is the momentum operator. It can be shown by an explicit calculation that
\eqn{15d} satisfies the quantum version of  \eqn{13a}:
\be
\ba{l}
{\displaystyle \left[ \hat{x}^a,  \hat{x}^b \right] =0,  }\\
\noalign {\vskip 1ex}
{\displaystyle  \left[ \hat{x}^a,  \hat{p}_{\displaystyle {}_b} \right] =i
\Pi_{ab}\!\left(\nabla_{\bf x }
\hat{f}
\right) = i \left( \delta_{ab} - { \partial_a \hat{f}  \partial_b\hat{f}  \over
\left(
\nabla_{\bf x } \hat{f}  \right)^2} \right),  } \\
\noalign {\vskip 1ex}
{\displaystyle \left[ \hat{p}_{\displaystyle {}_a},  \hat{p}_{\displaystyle
{}_b} \right] =
{i \over 2}
\left[ \left\{
\hat{p}_{\displaystyle {}_c} , {  \partial_a \partial_c \hat{f} \partial_b
\hat{f} \over
\left(
\nabla_{\bf x }
\hat{f}
\right)^2 }  \right\}  -  \left\{
\hat{p}_{\displaystyle {}_c} , {  \partial_b \partial_c \hat{f} \partial_a
\hat{f}  \over
\left(
\nabla_{\bf x }
\hat{f} \right)^2 } \right\} \right], }\ea \lab{16d}
\ee
where $\left\{\hat{A}, \hat{B} \right\} \equiv \hat{A} \hat{B} +
\hat{B}\hat{A}. $

\begin{itemize}

\item(iii) ${\cal O}\left(\hat{\bf p}, \hat{\bf x}\right) \equiv {\hat{\bf p}
}^2$:
with a similar manner as above, we find

\end{itemize}

\be
\ba{l}
{\displaystyle  \langle {\hat{\bf p} }^2 \rangle  =   \int
d^{D+1} \overline{\bf x}  \   \Pi_{ab}\!\left(\nabla_{\overline{\bf x}} f
\right)   \
\left( - { \partial^2 \over \partial \Delta x^a \partial \Delta x^b } \right)
\Bigg[ \delta
\!\left(
\overline{f}  \right)  { \left| \nabla_{\bf x} f({\bf x}) \cdot
\nabla f \right| \over \left( \nabla f \right)^2 }   }\\
\noalign {\vskip 1ex}
{\displaystyle  \qquad \qquad \qquad \times \varphi^*\! \left(\overline{\bf x}
+
{\Delta {\bf x} \over 2} \right)
\psi\! \left(\overline{\bf x} - {\Delta {\bf x} \over 2} \right) \Bigg]
\Bigg|_{\Delta {\bf x}=0 }   }\\
\noalign {\vskip 1ex}
{\displaystyle  =  \int d^{D+1} {\bf x}  \
  \delta \!\left( f ({\bf x} ) \right)  \varphi^*\! \left({\bf x} \right)
\Bigg[ - \Pi_{ab}\!\left(\nabla_{\bf x } f \right){ \partial^2
\over \partial
\Delta x^a \partial \Delta x^b }   }\\
\noalign {\vskip 1ex}
{\displaystyle + \left( \Pi_{ab}\!\left(\nabla_{\bf x } f \right)  {
\partial_a
\partial_c f \partial_c f  \over \left( \nabla_{\bf x } f \right)^2 } -
\partial_a
\Pi_{ab}\!\left(\nabla_{\bf x } f \right)  \right) { \partial \over \partial
x^b }  + {1
\over 2} \partial_b \left(
\Pi_{ab}\!\left(\nabla_{\bf x } f
\right) {  \partial_a
\partial_c f \partial_c f  \over \left( \nabla_{\bf x } f \right)^2 }  \right)
}\\
\noalign{\vskip1ex}
{\displaystyle \qquad \qquad  - {1 \over 4}  \partial_a \partial_b
\Pi_{ab}\!\left(\nabla_{\bf x } f
\right) - {1 \over 6} \Pi_{ab}\!\left(\nabla_{\bf x } f \right) {  \partial_a
\partial_b
 \partial_c f \partial_c f  \over \left( \nabla_{\bf x } f \right)^2 } \Bigg]
\psi\! \left(
{\bf x} \right) . }
 \ea
\lab{17d}
\ee
{}From this we obtain
\be
\ba{l}
{\displaystyle  {\hat{\bf p} }^2 = - \Pi_{ab}\!\left(\nabla_{\bf x } f \right){
\partial^2 \over \partial
\Delta x^a \partial \Delta x^b }   }\\
\noalign {\vskip 1ex}
{\displaystyle \qquad \ + \left\{  {  2\partial_a
\partial_b f \partial_a f  +  {\nabla_{\bf x }}^2 f \partial_b f \over \left(
\nabla_{\bf
x } f \right)^2 }  -  {  3 \partial_a
\partial_c f \partial_a f  \partial_b f \partial_c f   \over \left(
\nabla_{\bf x } f \right)^4 }  \right\}  {\partial \over \partial x^b} } \\
\noalign {\vskip 1ex}
{\displaystyle \qquad \ + { 1 \over \left( \nabla_{\bf x
} f \right)^2 }  \left\{ { 5 \over 6} \partial_a {\nabla_{\bf x }}^2 f
\partial_a f +  { 1 \over 4} \left( {\nabla_{\bf x }}^2 f \right)^2
+  { 3 \over 4} \partial_a \partial_b f  \partial_a \partial_b f \right\} }\\
\noalign {\vskip 1ex}
{\displaystyle \qquad \ - { 1 \over \left( \nabla_{\bf x
} f \right)^4 }  \Bigg\{ { 3 \over 2} {\nabla_{\bf x }}^2 f
\partial_a \partial_b f \partial_a f\partial_b f  +  { 7 \over 2} \partial_a
\partial_b
f \partial_b \partial_c f \partial_a f \partial_c f  }\\
\noalign {\vskip 1ex}
{\displaystyle \qquad  \qquad \qquad \qquad \qquad  \qquad \qquad +  { 5
\over 6}  \partial_a \partial_b \partial_c f  \partial_a f \partial_b f
\partial_c
f \Bigg\}  }\\
\noalign {\vskip 1ex}
{\displaystyle \qquad \  + { 4 \over \left( \nabla_{\bf x
} f \right)^6 }  \partial_a \partial_b f \partial_c \partial_d f \partial_a f
\partial_b f \partial_c
f \partial_d f .}
 \ea \lab{18d}
\ee

It should be noted that ${\hat{\bf p}}^2 \neq \hat{p}_{\displaystyle
{}_a}\hat{p}_{\displaystyle {}_a}$ unless $f({\bf x})$ is linear in ${\bf x}$.

\section{Discussion}

In this paper we have established a form of constraints in the path integral
formula
given by the time discretization. The main interest is how to incorporate the
classical constraint ${\bf p} \cdot \nabla_{\bf x} f =0$ into the quantum one:
the
correct form can be found by requiring that the delta
function be single-valued.

The conclusion is unchanged even if we take a nonstandard form of Hamiltonian
instead
of \eqn{8a} such as
\be
H( {\bf p} , {\bf x}) \longrightarrow h({\bf p}^2) + V({\bf x}),  \lab{1e}
\ee
provided $h'({\bf p}^2) \neq 0$.

Therefore we have successfully described a `local' form of the path integral
formula;
where the word `local' must be attached since if manifold is nontrivial and
composed of
$G/H$ there emerge induced gauge fields according to recent studies
\cite{LMT}\cite{OK}.  Our formula apparently lucks these informations. There
has been
a trial \cite{MT} but we are still on the way to the final goal.

\end{document}